\documentclass[preprint,pra,onecolumn]{revtex4}%
\usepackage{amsfonts}
\usepackage{amsmath}
\usepackage{amssymb}
\usepackage{graphicx}%
\providecommand{\U}[1]{\protect\rule{.1in}{.1in}}

\begin{document}
\title{Heralded entanglement of two distant quantum dot spins via optical interference}
\author{Li-Bo Chen$^{1}$}
\author{Wen Yang$^{2}$}
\email{wenyang@csrc.ac.cn}
\author{Zhang-Qi Yin$^{3}$}

\affiliation{$^{1}$School of Science, Qingdao Technological University, Qingdao 266033,
China }
\affiliation{$^{2}$Beijing Computational Science Research Center, No.3 Heqing Road Haidian
District Beijing 100084, China}
\affiliation{$^{3}$Center for Quantum Information, Institute for Interdisciplinary
Information Sciences, Tsinghua University, Beijing, 100084, People's Republic
of China}

\begin{abstract}
We present a proposal for heralded entanglement between two quantum dots via
Hong--Ou--Mandel effect. Each of the quantum dots, drived off-resonance by two
lasers, can be entangled with the coherent cavity mode. The output photons
from the two coherent cavity modes interfering by a beamsplitter, we could
entangle the two QDs with nearly unit success probability. Our scheme requires
neither direct coupling between qubits nor the detection of single photons.
Moreover the quantum dots do not need to have the same frequencies and
coupling constants.

\end{abstract}

\maketitle

\section{Introduction}

Quantum entanglement is treated as a crucial resource in many quantum
information tasks, such as quantum teleportation \cite{bennett}, quantum dense
coding \cite{bennett2}, quantum cryptography and quantum computation
\cite{Ekert}. Most of the tasks require generating entanglement among distant
quantum nodes. However, it is not easy to generate entanglement between
distant nodes, as interaction between qubit is generally local. In order to
solve the problem, many schemes of creating entanglement between spatially
separated nodes have been proposed
\cite{cirac,Enk,feng,yao1,yao2,yin,libo,Schwager1,Schwager2} and some
experimentally demonstrated \cite{duan,duan2,Sangouard,Bernien,Usmani}. These schemes are
either probabilistic or have yet to be demonstrated experimentally.

Recently, spin qubits in semiconductor quantum dots (QDs) attract much
interest because of their potential for a compact and scalable quantum
information architecture, and relatively long coherence time \cite{loss}. It
has been demonstrated that double quantum dots can be entangled by directly
coupling \cite{D.kim}. However, the method cannot be used to entangle the
distant quantum dots. It is found that semiconductor QDs can be strongly
coupled to photonic crystal cavity systems \cite{Hennessy,Englund}. However,
due to the dot size variation, semiconductor QDs usually have different
radiative properties, which make it difficult to generate entanglement by
indistinguishable emitting photons, as those have been done in atomic systems.
There are two approaches to overcome the problem: (i) tuning the QDs into
resonance by using externally applied strain \cite{Flagg,Patel}, or
controlling the Overhauser field \cite{x.xu2,gong} or (ii) making a detuning
between quantum dots and the same frequency output cavity modes
\cite{Busch,Sridharan,waks}, or using quantum frequency conversion
\cite{Ates,Zaske}. There has, to our best knowledge, been no reported
experimental realization of entanglement between two distant QDs. In Ref.
\cite{Busch} QDs in low Q cavities are coupled to a common high-Q cavity mode.
There are no reliable device structures like this yet. In
Ref.~\cite{Sridharan,waks} weak coherent fields being reflected from the
Cavity-QDs system have low efficiency or success possibility. Recently, schemes for
robust multiphoton entanglement creation using coherent (or Gaussian) state
light were put forward \cite{chan,cohen}. The rate is much larger than the
single-photon entanglement creation schemes \cite{duan,duan2,Sangouard,Usmani}.

In this paper, we describe a protocol for generating entanglement between two
quantum dots via Hong--Ou--Mandel effect \cite{hom}. The QDs are strongly
driven by $V$ polarized laser fields, generating $H$ polarized coherent cavity
modes which separately entangle with the two QDs. The output photons from the
two coherent cavity modes interfering by a beamsplitter, we could entangle the
two QDs with nearly unit efficiency and fidelity. The scheme does not need the
quantum dots to have the same radiation frequency. Compared with the
entanglement distributing schemes reported previously
\cite{duan,duan2,Sangouard}, our scheme combines both the advantages that
appear in direct coupling method (high efficiency) and single photon
interference method (high fidelity) \cite{Spiller}. We believe that our
protocol is a promising route to a scalable quantum computation in solid system.

\section{Entanglement of two quantum dots}

Suppose that two quantum dots are seperately coupled to two cavities with same
cavity mode frequency. The internal level configuration of each quantum dot is
shown in the left part of Fig.~1 \cite{Busch,x.xu,x.xu2}. The $\left\vert
X+\right\rangle _{i}\rightarrow\left\vert T+\right\rangle _{i}$ transition of
dot $i$ is driven by a $V$ polarized laser pulse with Rabbi frequency
$\Omega_{+}^{i}$ and detuning $\Delta_{+}^{i}$. Another $V$ polarized laser
pulse drives the transition $\left\vert X-\right\rangle _{i}\rightarrow
\left\vert T-\right\rangle _{i}$ with Rabbi frequency $\Omega_{-}^{i}$ and
detuning $\Delta_{-}^{i}$. The $\left\vert X+\right\rangle _{i}\rightarrow
\left\vert T-\right\rangle _{i}$\ and $\left\vert X-\right\rangle
_{i}\rightarrow\left\vert T+\right\rangle _{i}$\ transitions couple the $H$
polarized cavity mode $a_{i}$ with detuning $\Delta_{-}^{i}$ and $\Delta
_{+}^{i}$ and coupling strength $g_{-}^{i}$ and $g_{+}^{i}$. (By choosing the
detunings, the two quantum dots can coupled the same frequency cavities).
Introducing the rotating wave approximation and choosing the appropriate
interaction picture, the Hamiltonian is ($\hbar=1$)%

\begin{align}
H  &  =\sum_{i=1,2}[\Omega_{+}^{i}e^{-i\Delta_{+}^{i}}\left\vert
X+\right\rangle _{i}\left\langle T+\right\vert +\Omega_{-}^{i}e^{-i\Delta
_{-}^{i}}\left\vert X-\right\rangle _{i}\left\langle T-\right\vert \nonumber\\
&  +g_{+}^{i}e^{-i\Delta_{+}^{i}}\left\vert X-\right\rangle _{i}\left\langle
T+\right\vert a_{i}^{+}+g_{-}^{i}e^{-i\Delta_{-}^{i}}\left\vert
X+\right\rangle _{i}\left\langle T-\right\vert a_{i}^{+}+H.c.]\text{,}%
\end{align}
under the condition%

\begin{equation}
\Omega_{+}^{i}\text{, }\Omega_{-}^{i}\text{, }g_{-}^{i}\text{, }g_{+}^{i}%
\ll\Delta_{-}^{i}\text{, }\Delta_{+}^{i}\text{,}%
\end{equation}
and choose%
\begin{equation}
\frac{\Omega_{+}^{i}g_{+}^{i}}{\Delta_{+}^{i}}=\frac{\Omega_{-}^{i}g_{-}^{i}%
}{\Delta_{-}^{i}}=\lambda_{i}\text{,}%
\end{equation}
the Hamiltonian (1) simplifies to%
\begin{equation}
H_{\text{eff}}=\sum_{i=1,2}[\lambda_{i}(\left\vert X-\right\rangle
_{i}\left\langle X+\right\vert +\left\vert X+\right\rangle _{i}\left\langle
X-\right\vert )a_{i}^{+}+H.c.]\nonumber\\
=\sum_{i=1,2}\lambda_{i}\sigma_{y}^{i}\left(  a_{i}^{+}+a_{i}\right)
\end{equation}
here $\sigma_{y}^{i}=\left\vert X-\right\rangle _{i}\left\langle X+\right\vert
+\left\vert X+\right\rangle _{i}\left\langle X-\right\vert $, so$\ \left\vert
y+\right\rangle _{i}=\frac{1}{\sqrt{2}}\left(  \left\vert X-\right\rangle
_{i}+\left\vert X+\right\rangle _{i}\right)  $ and $\left\vert y-\right\rangle
_{i}=\frac{1}{\sqrt{2}}\left(  \left\vert X-\right\rangle _{i}-\left\vert
X+\right\rangle _{i}\right)  $.%
\begin{equation}
e^{-iH_{\text{eff}}t}\left\vert y+\right\rangle _{i}\left\vert 0\right\rangle
_{c}^{i}=\left\vert y+\right\rangle _{i}\left\vert -\alpha_{i}\right\rangle
_{c}^{i}\text{,}\nonumber
\end{equation}%
\begin{equation}
e^{-iH_{\text{eff}}^{\prime}t}\left\vert y-\right\rangle _{i}\left\vert
0\right\rangle _{c}^{i}=\left\vert y-\right\rangle _{i}\left\vert \alpha
_{i}\right\rangle _{c}^{i}\text{,}\nonumber
\end{equation}
here $\alpha_{i}=-i\lambda_{i}t$, and $\left\vert m\right\rangle _{c}^{i}$
represents the state of the cavity mode $a_{i}$. If our initial state of the
dot--cavity combined system is $\sum_{i=1,2}\left\vert X-\right\rangle
_{i}\left\vert 0\right\rangle _{c}^{i}$, under the Hamiltonian $H_{\text{eff}%
}$, the state of the system evolve to%

\begin{equation}
\left\vert \psi\left(  t\right)  \right\rangle =e^{-iH_{\text{eff}}t}%
{\displaystyle\prod\limits_{i=1,2}}
\left\vert X-\right\rangle _{i}\left\vert 0\right\rangle _{c}^{i}\nonumber\\
=\frac{1}{2}e^{-iH_{\text{eff}}t}%
{\displaystyle\prod\limits_{i=1,2}}
\left(  \left\vert y+\right\rangle _{i}+\left\vert y-\right\rangle
_{i}\right)  \left\vert 0\right\rangle _{c}^{i}\nonumber\\
=\frac{1}{2}%
{\displaystyle\prod\limits_{i=1,2}}
\left(  \left\vert y+\right\rangle _{i}\left\vert \alpha_{i}\right\rangle
_{c}^{i}+\left\vert y-\right\rangle _{i}\left\vert -\alpha_{i}\right\rangle
_{c}^{i}\right)  \text{.}%
\end{equation}
The mode $\widehat{a}_{1}^{out}$ ( $\widehat{a}_{2}^{out}$ ) output from the
cavity $1$ ($2$) reaches the beamsplitter ($BS$). The $BS$ then applies the transformations%

\begin{align}
\widehat{a}_{1}^{out}  &  \rightarrow\frac{\widehat{c}+\widehat{d}}{\sqrt{2}%
}\text{,}\\
\widehat{a}_{2}^{out}  &  \rightarrow\frac{\widehat{c}-\widehat{d}}{\sqrt{2}%
}\text{,}%
\end{align}
where $\widehat{c}\ $and $\widehat{d}$ are the bosonic modes monitored by
detector $c$ and detector $d$ respectively. In the case of $\alpha_{1}%
=\alpha_{2}=\alpha$, the state of the system becomes%

\begin{align}
&  \left\vert \psi\left(  t\right)  \right\rangle _{f}=\frac{1}{2}(\left\vert
y+\right\rangle _{1}\left\vert y+\right\rangle _{2}\left\vert \sqrt{2}%
\alpha\right\rangle _{c}\left\vert \text{vac}\right\rangle _{d}+\left\vert
y+\right\rangle _{1}\left\vert y-\right\rangle _{2}\left\vert \text{vac}%
\right\rangle _{c}\left\vert -\sqrt{2}\alpha\right\rangle _{d}\nonumber\\
&  \text{ \ \ \ \ }+\left\vert y-\right\rangle _{1}\left\vert y+\right\rangle
_{2}\left\vert \text{vac}\right\rangle _{c}\left\vert \sqrt{2}\alpha
\right\rangle _{d}+\left\vert y-\right\rangle _{1}\left\vert y-\right\rangle
_{2}\left\vert -\sqrt{2}\alpha\right\rangle _{c}\left\vert \text{vac}%
\right\rangle _{d})\text{,}%
\end{align}
where The state $\left\vert \text{vac}\right\rangle $ represents the vacuum
state of field mode. Conditioned on photons detection event at detector $c$,
the state of the QDs collapses onto%

\begin{equation}
\left\vert \phi\right\rangle _{1}=\frac{1}{\sqrt{2}}(\left\vert
y+\right\rangle _{1}\left\vert y+\right\rangle _{2}+\left\vert y-\right\rangle
_{1}\left\vert y-\right\rangle _{2})\text{,}%
\end{equation}
photons detection event at detector $d$, the state of the QDs collapses onto%
\begin{equation}
\left\vert \phi\right\rangle _{2}=\frac{1}{\sqrt{2}}(\left\vert
y+\right\rangle _{1}\left\vert y-\right\rangle _{2}+\left\vert y-\right\rangle
_{1}\left\vert y+\right\rangle _{2})\text{,}%
\end{equation}
It should be noted that, so long as detector is triggered, a perfect entangled
state can be generated heraldedly even when the QDs have different resonant frequencies.

\section{Simulations with the Lindblad master equation}

The system consists of two cavity modes $\hat{a}_{1},\hat{a}_{2}$ and two
quantum dots. In the absence of any photon detection, the system evolution
obeys the Lindblad master equation%

\begin{align}
\dot{\rho}(t)  &  =i[H,\rho(t)]-\sum_{i=1,2}[\frac{\kappa_{i}}{2}\left(
a_{i}^{+}a_{i}\rho(t)-2a_{i}^{+}\rho(t)a_{i}+\rho(t)a_{i}^{+}a_{i}\right)
\nonumber\\
&  -\frac{\Gamma_{i}}{2}\left(  \left\vert X+\right\rangle _{i}\left\langle
X+\right\vert \rho(t)-2\left\vert X-\right\rangle _{i}\left\langle
X+\right\vert \rho(t)\left\vert X+\right\rangle _{i}\left\langle X-\right\vert
+\rho(t)\left\vert X+\right\rangle _{i}\left\langle X+\right\vert \right)
]\text{,}%
\end{align}
where $\Gamma_{i}$ denotes the relaxation of the two QDs and $\kappa_{i}$ the
damping of the two cavities. The steady state $\hat{\rho}_{\mathrm{ss}}(t)$ of
the system can be obtained by solving the Lindblad master equation.

Now each cavity is coupled to a transmission line. The output of these two
transmission lines are mixed by a $50:50$ beam splitter and each port of the
beamsplitter is monitored by photon counter. The output electric field
$\hat{b}_{j,\mathrm{out}}(t)$ of the $j$th $(j=1,2)$ transmission line carries
information about the cavity:%

\begin{align}
\hat{b}_{1,\mathrm{out}}(t)  &  =-i\sqrt{\kappa_{1}}\hat{a}_{1}(t)\text{,}%
\nonumber\\
\hat{b}_{2,\mathrm{out}}(t)  &  =-i\sqrt{\kappa_{2}}\hat{a}_{2}(t)\text{,}%
\end{align}
where the vacuum fluctuation part $\hat{b}_{1,\mathrm{in}}(t)$ and $\hat
{b}_{2,\mathrm{in}}(t)$ have been neglected, since they have no effect on the
photon counter. The output field of the beam splitter is%

\begin{align}
\hat{c}(t)  &  =[\hat{b}_{1,\mathrm{out}}(t)+i\hat{b}_{2,\mathrm{out}%
}(t)]/\sqrt{2}\nonumber\\
&  =-i[\sqrt{\kappa_{1}}\hat{a}_{1}(t)+\sqrt{\kappa_{2}}\hat{a}_{2}%
(t)]/\sqrt{2}\text{,}%
\end{align}

\begin{align}
\hat{d}(t)  &  =[\hat{b}_{1,\mathrm{out}}(t)-i\hat{b}_{2,\mathrm{out}%
}(t)]/\sqrt{2}\nonumber\\
&  =-i[\sqrt{\kappa_{1}}\hat{a}_{1}(t)-\sqrt{\kappa_{2}}\hat{a}_{2}%
(t)]/\sqrt{2}\text{.}%
\end{align}
During $[t,t+dt]$ ($dt\rightarrow0$ is an infinitesimal interval), photon
detection $c$ occurs with probability%
\begin{equation}
p\left(  dt\right)  =dt\operatorname*{Tr}\hat{c}\hat{\rho}_{\mathrm{ss}%
}(t)\hat{c}^{\dagger}\text{,}%
\end{equation}
after which the system collapses to the state (un-normalized) $\tilde{\rho
}^{(1)}(t+dt)=\hat{c}\hat{\rho}_{\mathrm{ss}}(t)\hat{c}^{\dagger}dt$. Fig.~2
plots of the average detecting photon number $N$ vs $\lambda t$ for different
values of $\kappa/\lambda$, here we choose $\lambda_{1}=\lambda_{2}=\lambda$,
$\kappa_{1}=\kappa_{2}=\kappa$, $\Gamma_{1}=\Gamma_{2}=\Gamma$, and
$\Gamma=0.05\lambda$. In the case of the detector $c$ (or $d$) is triggered,
Fig.~2 plots the fidelity of the entanglemnt state $F$ vs $\lambda t$ for
different values of $\kappa/\lambda$, here we choose $\kappa=\lambda$.

We use parameters appropriate for self-assembled InAs QDs, $\delta=\Delta
_{-}-\Delta_{+}=(\omega_{T-}-\omega_{T+})+(\omega_{X-}-\omega_{X+})=46$ $\mu
$eV \cite{x.xu,Economou}, $g_{+}=g_{-}=90$ $\mu$eV \cite{Hennessy},
$\Delta_{-}=460$ $\mu$eV, $\Delta_{+}=414$ $\mu$eV, $\Omega_{+}=41.4$ $\mu$eV,
and $\Omega_{-}=46$ $\mu$eV. With these parameters, the conditions given by
equations (2) and (3) are fulfilled and value $\lambda=2.2$ $G$Hz is obtained.
In realistic experiments, the spontaneous decay rate of the trion state
$\left\vert T\pm\right\rangle $ is about $\Gamma_{T}=2\pi\times130$ MHz
\cite{x.xu3}. Fig.~4 plots the population of the trion state $\left\vert
T\pm\right\rangle $ by using the above parameters. Due to large detuning, the
effective decay rate could be estimated as $\Gamma_{T}\Omega g/\Delta
^{2}\approx2\pi\times2.6$ MHz \cite{Pellizzari,W.L.Yang}, which is much
smaller than $\lambda$. This implies that the influence from the spontaneous
decay in our scheme. The main source of error is the depahsing of elecron
spin. The typical dephasing rate of the InAs QD electron spin has been
measured to be about $\Gamma=\left(  10ns\right)  ^{-1}\approx0.05\lambda$
\cite{Bracker}, the fidelity of the entanglemnt state is larger than $0.95$ as
shown in Fig. 3. From the Fig. 3. and 4 we can neglect the influence the trion
state $\left\vert T\pm\right\rangle $.

\section{Conclusion}

In conclusion, we have generated deterministic entanglement between two
distant quantum dots using classical interference; that is, there is no
inherent probabilistic nature to our quantum entangling source. The method is
robust to the difference of the two quantum dots and cavities. Our scheme does
not need to control coupling between qubits nor to detection of single
photons. Using this larger arrays of quantum dot qubits could be linked
together for scale-up to a quantum computer \cite{Stoneham}.

\textbf{Acknowledgments: }We thanks Prof. Lu Jeu Sham and Dr. Jianqi Zhang for helpful discussions. This research is supported the National Natural Science Foundation of China (Grant No.11304174), Natural
Science Foundation of Shandong Province (Grant No. ZR2013AQ010), and the
research starting foundation of Qingdao Technological University.

\begin{figure}[ptbh]
\includegraphics[scale=1, angle=0]{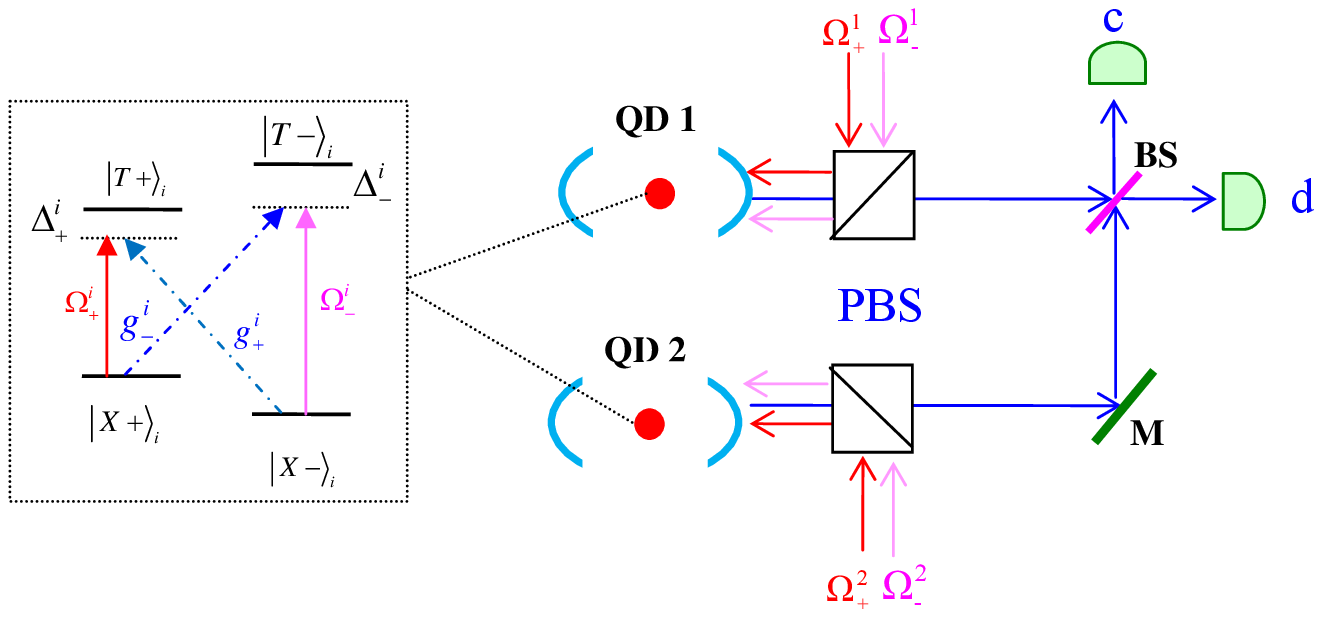}\caption{[Color online] Two
quantum dots are embedded in two cavities with the same frequency. The
$\left\vert X+\right\rangle _{i}\rightarrow\left\vert T+\right\rangle _{i}$
transition of dot $i$ is driven by a $V$ polarized laser pulse with Rabbi
frequency $\Omega_{+}^{i}$ and detuning $\Delta_{+}^{i}$. Another $V$
polarized laser pulse drives the transition $\left\vert X-\right\rangle
_{i}\rightarrow\left\vert T-\right\rangle _{i}$ with Rabbi frequency
$\Omega_{-}^{i}$ and detuning $\Delta_{-}^{i}$. The $\left\vert
X+\right\rangle _{i}\rightarrow\left\vert T-\right\rangle _{i}$\ and
$\left\vert X-\right\rangle _{i}\rightarrow\left\vert T+\right\rangle _{i}%
$\ transitions couple the $H$ polarized cavity mode $a_{i}$ with detuning
$\Delta_{-}^{i}$ and $\Delta_{+}^{i}$ and coupling strength $g_{-}^{i}$ and
$g_{+}^{i}$. After passing though Polarized Beam Splitters ($PBS$s), the
output modes are mixed with a $50:50$ Beam Splitter ($BS$) and measured by
photodetectors $c$ and $d $.}%
\end{figure}

\newpage\begin{figure}[ptbh]
\includegraphics[scale=1, angle=0]{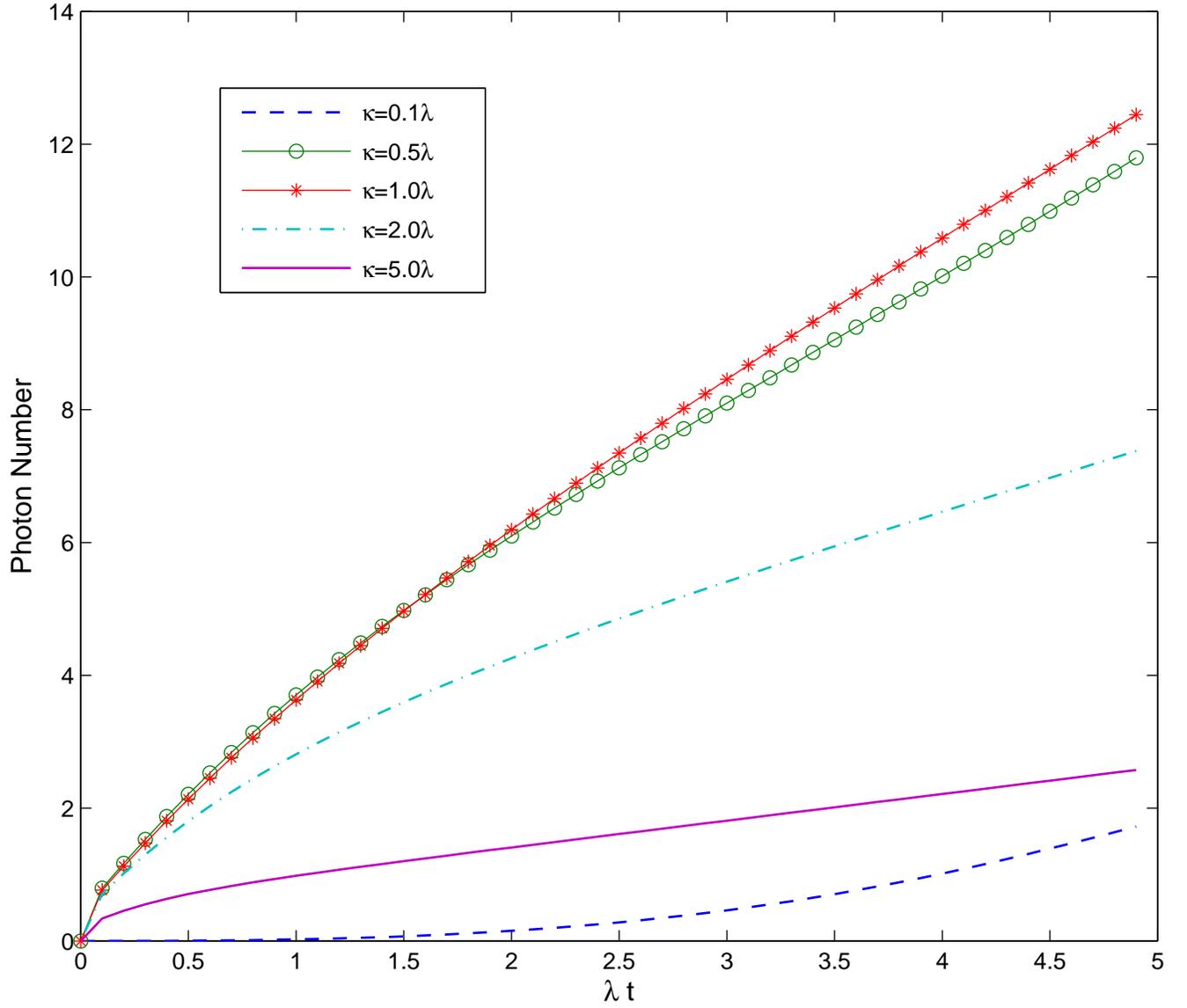}\caption{[Color online] The
average leaking photon number $N$ vs $\lambda t$ for different values of
$\kappa/\lambda$ (From bottom to top, $\kappa/\lambda=0.1$, $0.2$, $0.5$,
$1.0$).}%
\end{figure}

\newpage\begin{figure}[ptbh]
\includegraphics[scale=1, angle=0]{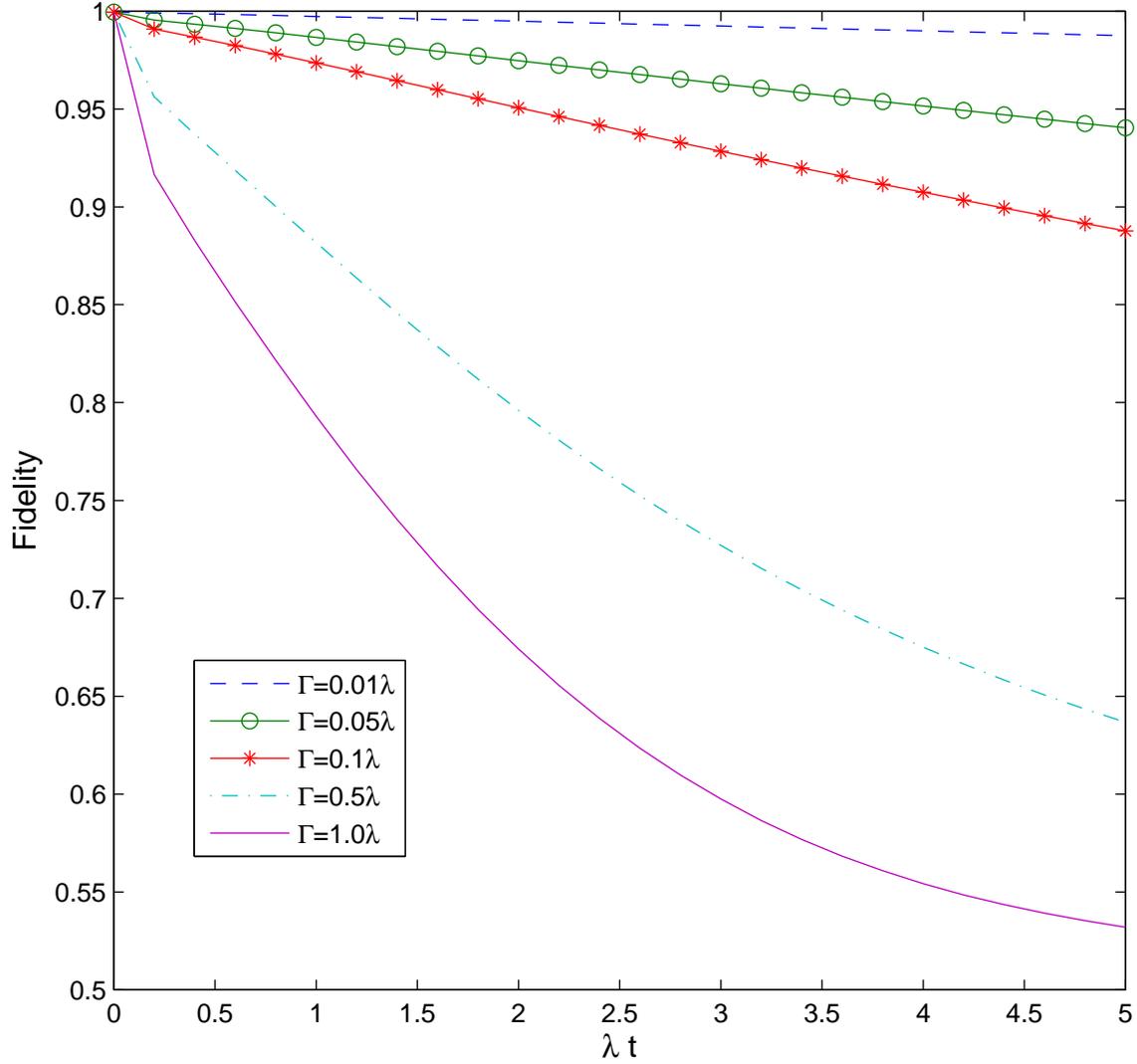}
\caption{The fidelity of the entanglemnt state $F$ vs $%
\protect\lambda t$ for different values of $\protect\kappa /\protect\lambda $
(From top to bottom, $\Gamma /\protect\lambda =0.01$, $0.05$, $0.1$, $0.5$, $%
1.0$), here we choose $\protect\kappa =\protect\lambda $.}
\end{figure}

\newpage\begin{figure}[ptbh]
\includegraphics[scale=1, angle=0]{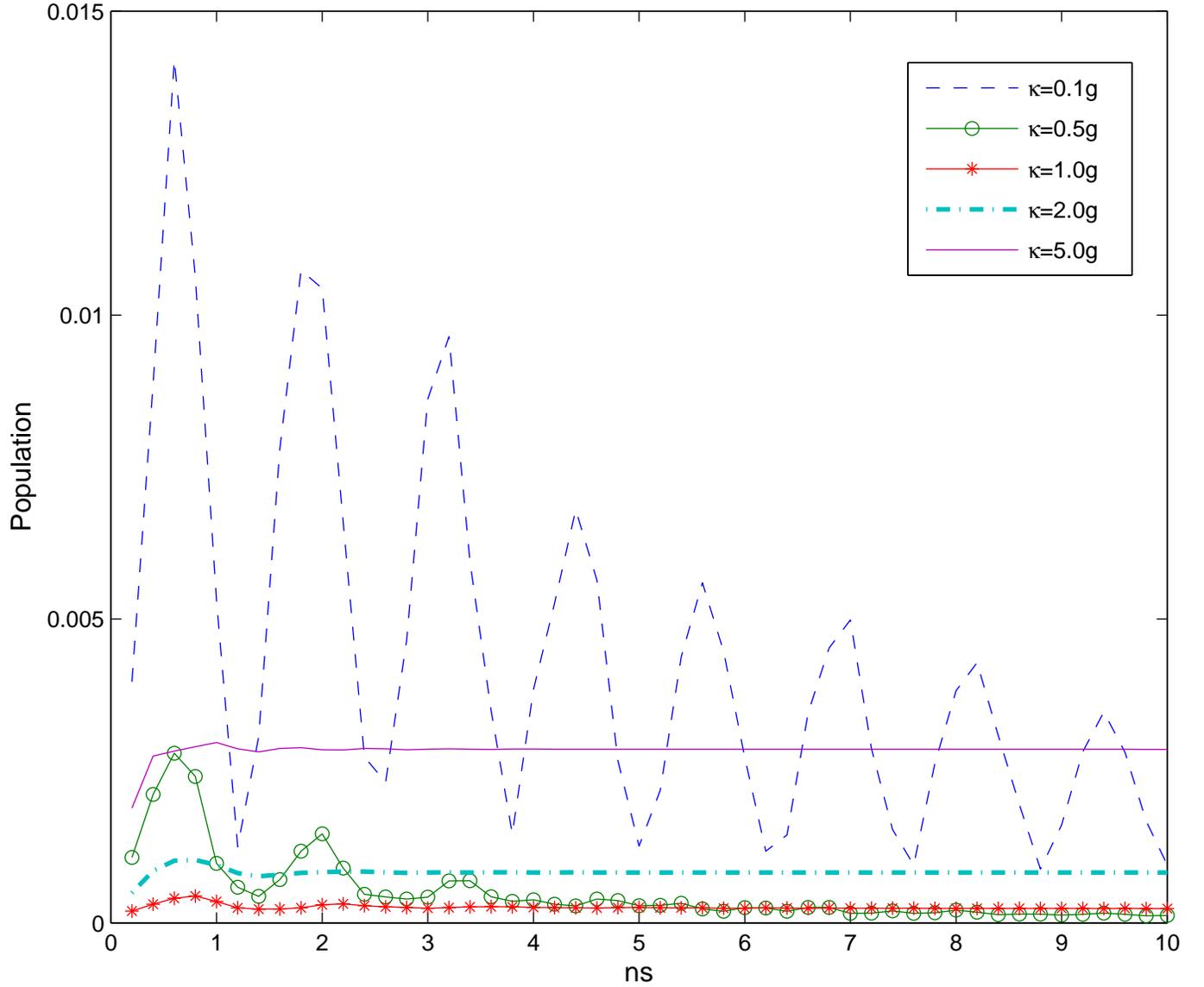}
\caption{The
population of the trion state $\left\vert T\pm \right\rangle $,
here we choose $\delta =46$
$\mu $eV , $g_{+}=g_{-}=90$ $\mu $eV , $%
\Delta _{-}=460$ $\mu $eV, $\Delta _{+}=414$ $\mu $eV, $\Omega _{+}=41.4$ $%
\mu $eV, and $\Omega _{-}=46$ $\mu $eV, $\Gamma _{T}=2\pi \times 130$ MHz.}
\end{figure}

\end{document}